\documentclass[onecolumn,aps,prl,groupedaddress,preprintnumbers,amsmath,amssymb]{revtex4}

\usepackage{graphicx}
\usepackage{here}

\begin{document}

\title{Single-photon quantum regime of artificial radiation pressure \\ on a surface acoustic wave resonator}

\author{Atsushi~Noguchi$^{1,2,3}$}
\email[]{u-atsushi@g.ecc.u-tokyo.ac.jp}
\author{Rekishu~Yamazaki$^{1}$}
\author{Yutaka~Tabuchi$^{1}$}
\author{Yasunobu~Nakamura$^{1,4}$}

\affiliation{%
$^{1}$Research Center for Advanced Science and Technology (RCAST), The University of Tokyo, Meguro-ku, Tokyo, 153-8904, Japan,\\
$^{2}$PRESTO, Japan Science and Technology Agency, Kawaguchi-shi, Saitama 332-0012, Japan,\\
$^{3}$Komaba Institute for Science (KIS), The University of Tokyo, Meguro-ku, Tokyo, 153-8902, Japan,\\
$^{4}$Center for Emergent Matter Science (CEMS), RIKEN, Wako-shi, Saitama 351-0198, Japan
}

\date{\today}

\begin{abstract}
Electromagnetic fields carry momentum, which upon reflection on matter gives rise to the radiation pressure of photons.
The radiation pressure has recently been utilized in cavity optomechanics for controlling mechanical motions of macroscopic objects at the quantum limit.
However, because of the weakness of the interaction, attempts so far had to use a strong coherent drive to reach the quantum limit
Therefore, the single-photon quantum regime, where even the presence of a totally off-resonant single photon alters the quantum state of the mechanical mode significantly, is one of the next milestones in cavity optomechanics.
Here we demonstrate an artificial realization of the radiation pressure of microwave photons acting on phonons in a surface acoustic wave resonator. 
The order-of-magnitude enhancement of the interaction strength originates in the well-tailored strong second-order nonlinearity of a superconducting Josephson-junction circuit.
The synthetic radiation pressure interaction adds a key element to the quantum optomechanical toolbox and can be applied to quantum information interfaces between electromagnetic and mechanical degrees of freedom.  
\end{abstract}

\maketitle


The radiation pressure of electromagnetic field~\cite{Nichols} is one of the fundamental concepts in cavity optomechanics~\cite{Kippenberg2014}.
Even though the interaction is rather weak at the single-photon level, one can apply a strong drive field to enhance the effective coupling strength to reach the quantum regime~\cite{Teufel2011b,Teufel2015c, noguchiMEM, Chan2011a, aspelmeyer, Purdy}.
Based on this interaction, ground-state cooling and quantum state control of mechanical oscillators have been reported on suspended membranes~\cite{Teufel2011b,Teufel2015c, noguchiMEM}, phononic-crystal cavities~\cite{Chan2011a, aspelmeyer, Purdy}, micro-toroidal resonators~\cite{Kippenberg2012}, and bulk oscillators~\cite{Sill2017}.
For such experiments, however, strong drive fields often restrict quantum-limited functionalities of the optomechanical systems by introducing noise, heat and other dissipations.

The single-photon quantum regime is reached when the radiation pressure of a single photon is strong enough to overcome other dissipations in the system~\cite{Kippenberg2014}, where the quantum state of the mechanical mode is coherently controlled by the quantum of the  electromagnetic field. 
However, such a strong radiation pressure interaction has been elusive in optomechanical systems studied so far, while there are a few experiments approaching this regime~\cite{Sillanpaa2015,Painter2014}.

\begin{figure*}[t]
\begin{center}
   \includegraphics[width=16.0cm,angle=0]{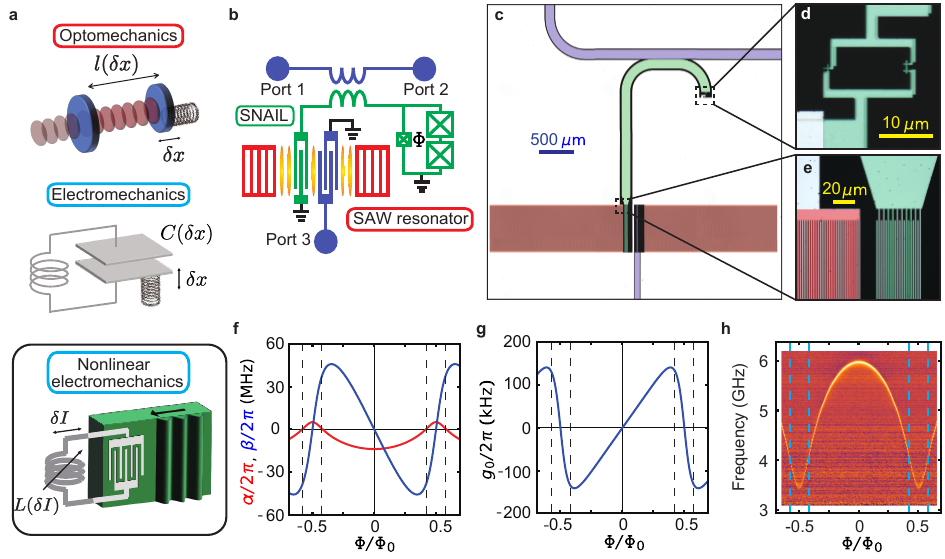}
\caption{
{\bf SAW-MW hybrid system for artificial radiation pressure interaction.}
{\bf{a.}}~Variety of optomechanical systems.
(top) Optomechanical system using an optical cavity.
The effective cavity length $l$ is modulated by the displacement of the mechanical oscillator $\delta x$.
(middle) Electromechanics using a microwave lumped-element circuit.
The capacitance $C$ is modulated by the displacement of the mechanical oscillator.
(bottom) Electromechanics with a nonlinear Josephson-junction circuit and a SAW resonator.
The inductance $L$ is modulated by the current induced by transduction from acoustic waves.
{\bf{b.}}~Schematic of the SAW optomechanical system.
A SAW resonator defined by Bragg mirrors~(red) couples to a nonlinear MW resonator (green) via an interdigitated transducer. 
A SNAIL loop consisting of three Josephson junctions works as a nonlinear inductive element.
Ports 1 and 2 are external feed lines for the MW resonator, and port~3 is that for the SAW resonator having a spatial mode shown in yellow.
{\bf{c}}--{\bf{e.}}~False-colored micrographs of the sample. The colors of the electrodes correspond to the ones in the schematic in {\bf b}.
{\bf{d.}}~Magnification of the SNAIL part. The three junctions form the SNAIL loop. 
{\bf{e.}}~Zoom-up of a part of the SAW resonator and the interdigitated transducer.
{\bf{f.}}~Calculated nonlinearity of the MW resonator as a function of the magnetic flux $\Phi$ penetrating through the SNAIL loop. 
Red (blue) curve represents the self-Kerr (Pockels) nonlinearity~$\alpha$~($\beta$) of the MW resonator.
{\bf{g.}}~Calculated strength $g_0$ of the artificial radiation pressure interaction induced by the nonlinearity of the SNAIL.
{\bf{h.}}~Spectrum of the nonlinear resonator as a function of $\Phi$ measured with a weak MW probe whose average intra-resonator photon number is much less than unity. 
Vertical dashed lines in {\bf{f}}--{\bf{h}} indicate flux bias conditions where the self-Kerr nonlinearity vanishes in the numerical simulation.
}
\label{fig1}
\end{center}
\end{figure*}

Here we introduce an artificial optomechanical system consisting of a surface acoustic wave (SAW) resonator and a superconducting circuit. 
The conventional radiation pressure arises from the frequency shift of the optical (or electrical) resonator depending on the displacement of the mechanical system~(Fig.~\ref{fig1}a).
Instead, we utilize a superconducting circuit with Josephson junctions, which are known to be a versatile platform for engineering strong nonlinearity with negligible dissipation~\cite{nakamura1999,walraff2004nat, Yamamoto_JPA, devoret2016}. 
The current induced by the acoustic waves in a piezoelectric material modulates the inductive energy of the Josephson circuit, which results in the motion-dependent frequency shift of the electrical resonator.
The enhanced artificial radiation pressure enables us to reach the single-photon quantum regime.


\section*{Results}
{\bf System and the model.}
Our hybrid system is composed of a Fabry-P\'{e}rot-type SAW resonator defined by a pair of Bragg mirrors~\cite{noguchiSAW}, and a nonlinear microwave~(MW) resonator~(Fig.~\ref{fig1}).
They are coupled to each other via an interdigitated transducer (IDT) through the piezoelectric interaction.
All the structures are made of aluminum evaporated on a ST-X quartz substrate (See the details in the Method and Supplementary Note 1~\cite{supple}).

The nonlinear MW resonator consists of a short coplanar waveguide connected to the IDT on one end.
On the other end it is grounded via a loop interrupted by one small and two large Josephson junctions with the Josephson energies $E_\mathrm{J}'$ and $E_\mathrm{J}$, respectively. 
The circuit element is called the Superconducting Nonlinear Asymmetric Inductive eLement (SNAIL)~\cite{SNAIL}.
The SNAIL has the inductive energy 
\begin{equation}
U(\theta )=-E_\mathrm{J}^\prime\cos \theta -2E_\mathrm{J}\cos \left( \frac{\phi-\theta}{2}\right) ,
\end{equation}
where $\theta$ is the superconducting phase across the small junction, $\phi =2\pi\Phi /\Phi _0$ is the reduced magnetic flux, $\Phi$ is the flux threading the loop, and $\Phi _0=h/2e$ is the flux quantum.
The SNAIL capacitively shunted with the coplanar waveguide forms the nonlinear MW resonator, whose Hamiltonian reads (with $\hbar =1$)
\begin{equation}
\hat{H}_\mathrm{m}=\omega _\mathrm{m}\hat{a}^\dagger\hat{a}+\alpha _0\hat{a}^\dagger\hat{a}^\dagger\hat{a}\hat{a}+
\beta (\hat{a}^\dagger\hat{a}^\dagger\hat{a} + \mathrm{h.c.} ).
\end{equation}
Here $\hat{a}$ ($\hat{a}^\dagger$) is the annihilation (creation) operator of a photon in the MW resonator, and $\omega _\mathrm{m}$ is the resonance frequency.
The terms with coefficients $\alpha_0$ and $\beta $ represent the nonlinearities of the resonator corresponding to the self-Kerr and Pockels effects, respectively (See the details in the Supplementary Note 2~\cite{supple}).

\begin{figure}[t]
\begin{center}
   \includegraphics[width=7.5cm,angle=0]{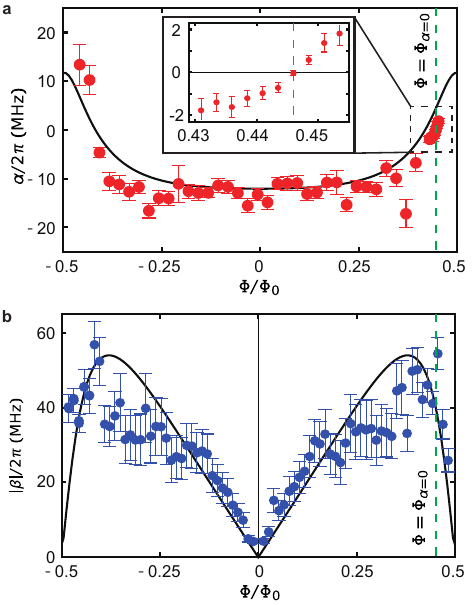}
\caption{
{\bf Nonlinearity of the MW resonator.}
{\bf a.}~Self-Kerr nonlinearity $\alpha$ as a function of the flux through the SNAIL loop.
Inset shows an enlarged view around $\Phi = \Phi_{\alpha = 0}$.
{\bf{b.}}~Absolute value of the Pockels nonlinearity $|\beta|$, obtained from the phonon-to-photon conversion experiment, as a function of the magnetic flux bias.
The curve in each panel shows the result of the numerical simulation without any fitting parameters.
Green dashed lines indicate $\Phi = 0.445\Phi _0\equiv \Phi_{\alpha =0}$.
The theoretical curves are the same as those in Fig.~1f. 
Error bars represent standard errors.
}
\label{fig2}
\end{center}
\end{figure}

The Hamiltonian of the hybrid system consisting of the MW and SAW resonators is written as
\begin{eqnarray}
\hat{H}=\hat{H}_0+\hat{V},
\end{eqnarray}
where
\begin{equation}
\hat{H}_0=\omega _\mathrm{m}\hat{a}^\dagger\hat{a}+\omega_\mathrm{s}\hat{b}^\dagger\hat{b},
\end{equation}
and
\begin{eqnarray}
\hat{V}=\alpha _0\hat{a}^\dagger\hat{a}^\dagger\hat{a}\hat{a}
+\beta (\hat{a}^\dagger\hat{a}^\dagger\hat{a} + \mathrm{h.c.} )+g_\mathrm{p}(\hat{a}^\dagger\hat{b}+\hat{a}\hat{b}^\dagger).
\end{eqnarray}
Here $\hat{b}$ $(\hat{b}^\dagger)$ is the annihilation (creation) operator of a phonon in the SAW resonator, $\omega _\mathrm{s}$ is the resonance frequency, and $g_\mathrm{p}$ is the piezoelectric coupling strength between the SAW and MW resonators.
By treating $\hat{V}$ as a perturbation (See the details in the Supplementary Note 3~\cite{supple}), 
we obtain an effective interaction Hamiltonian
\begin{eqnarray}
\hat{V}_\mathrm{eff}&=&\left( \alpha _0 -\frac{3\beta ^2}{\omega _\mathrm{m}}\right) \hat{a}^\dagger\hat{a}^\dagger\hat{a}\hat{a}
-\left( \frac{2g_\mathrm{p}\beta}{\omega _\mathrm{m}-\omega _\mathrm{s}}\right) \hat{a}^\dagger\hat{a} (\hat{b}^\dagger +\hat{b}),\nonumber\\
&\equiv &\alpha \hat{a}^\dagger\hat{a}^\dagger\hat{a}\hat{a}+g_0 \hat{a}^\dagger\hat{a} (\hat{b}^\dagger +\hat{b}),
\end{eqnarray}
under the rotating-wave approximation. 
This derivation is valid when $\{\omega _\mathrm{m},~\omega_\mathrm{s},~\omega_\mathrm{m}-\omega_\mathrm{s}\}\gg \{ |\alpha _0| ,~|\beta |,~|g_\mathrm{p}|\}$ and $\omega _\mathrm{m}\gg\omega _{\mathrm{s}}$ are satisfied.
The second term on the right-hand side represents an artificial radiation pressure interaction analogous to the Pockels effect.
On the other hand, the undesired first term corresponds to a self-Kerr nonlinearity, which can be eliminated by finding experimental conditions where $\alpha$ vanishes.
As we will show later, this condition mitigates the saturation effect and provides full functionality of the realized artificial radiation pressure.

Figure \ref{fig1}f shows the calculated strengths of the self-Kerr nonlinearity $\alpha$ and Pockels nonlinearity $\beta $ for the parameters of our sample.
Notably, $\alpha$ vanishes at certain flux bias conditions $\Phi = \{ \Phi_{\alpha=0}, \Phi_0 - \Phi_{\alpha=0} \}\! \mod \Phi_0$ (vertical dashed lines in Figs.~1f--h), while $\beta$ remains finite at the conditions.
In canonical optomechanical systems, the resonance frequency of the optical (or electrical) resonator is directly affected by the displacement of the mechanical oscillator.
Here, in contrast, the resonance frequency of the MW resonator is modulated by the current excited by the mechanical oscillations through the piezoelectric effect, resulting in the Pockels nonlinearity and the synthetic optomechanical coupling.
Figure \ref{fig1}g shows the calculated strength $g_0$ of the artificial radiation pressure interaction. 
It is of importance that $g_0$ takes a large value at the flux bias condition where $\alpha$ vanishes.

\vspace{3mm}

{\bf Nonlinearity of the circuit.}
Figure \ref{fig2}a shows the experimentally-determined self-Kerr nonlinearity $\alpha$ as a function of the flux bias.
For that, we measure the shift of the MW resonator frequency per average intra-resonator probe photon number as a function of the flux bias.
The observed self-Kerr nonlinearity changes its sign near $\Phi = \pm 0.5 \Phi_0$, as expected.
The relatively large scattering of the experimental data points are presumably due to the uncertainty in the determination of the probe photon number in the resonator because of the strongly flux-dependent loss rates of the MW resonator (See Supplementary Figure 1 in the Supplementary Note 1~\cite{supple}). 
As shown in the inset, $\alpha$ vanishes at $\Phi _{\alpha =0} \equiv 0.445\Phi _0$.
At this bias point, the MW resonator frequency is $\omega_\mathrm{m}/2\pi = 3.85$~GHz~(Fig.~1h), largely detuned from the SAW resonator frequency $\omega_\mathrm{s}/2\pi = 785.25$~MHz.

\begin{figure}[ht]
\begin{center}
   \includegraphics[width=7.5cm,angle=0]{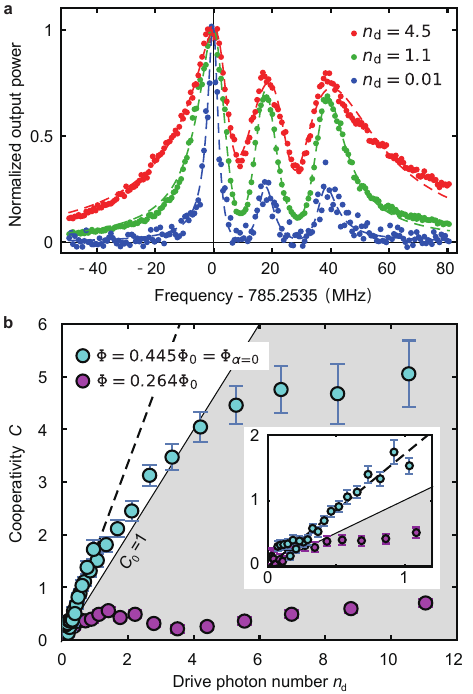}
\caption{
{\bf Strong artificial radiation pressure.}
{\bf{a.}}~Normalized output power of the MW signal coherently up-converted from the SAW excitations as a function of the detuning of the SAW drive frequency (dots). 
Three datasets are for different drive powers represented by the average number, $n_\mathrm{d}$, of drive photons in the MW resonator.
Dashed curves are the results of fittings with a sum of three Lorentzians in the complex plane.
The left-most peak is from the fundamental transverse mode of the SAW resonator, 
while the two other peaks are due to higher-index transverse modes.
{\bf{b.}}~Cooperativity $C$ as a function of the drive photon number $n_\mathrm{d}$.
White area corresponds to the single-photon quantum regime, i.e., $C_0>1$.
Cyan dots are the experimental data at $\Phi = 0.445 \Phi_0 = \Phi_{\alpha=0}$.
Purple dots are taken at $\Phi = 0.264 \Phi_0$, where $\alpha \neq 0$ and the saturation takes place at lower power.
Black dashed line is the linear fit for the cyan dots in the low-power region.
Inset shows an enlarged view near the origin.
Error bars represent standard errors.
}
\label{fig3}
\end{center}
\end{figure}

To evaluate the strength of Pockels nonlinearity $\beta$, we perform a phonon-to-photon conversion experiment from the SAW resonator to the MW resonator.
We irradiate the MW resonator with the red-sideband drive at frequency $\omega _\mathrm{d}\sim \omega _\mathrm{m}-\omega _\mathrm{s}$.
Under the rotating wave approximation in the resolved-sideband limit , i.e., $\kappa \ll \omega_\mathrm{s}$, where $\kappa$ is the total decay of the MW resonator, the Hamiltonian becomes,
\begin{eqnarray}
\hat{H}&=& \Delta\hat{a}^\dagger\hat{a}+g_0\sqrt{n_\mathrm{d}}(\hat{a}^\dagger\hat{b}+\hat{a}\hat{b}^\dagger ),
\end{eqnarray}
where $n_\mathrm{d}$ is the average photon number of the drive field in the MW resonator and $\Delta \equiv \omega_\mathrm{m}-\omega_\mathrm{d}-\omega _\mathrm{s}$ is the detuning.
The details of the derivation are presented in the Supplementary Note 4~\cite{supple}.
When the drive field is tuned to the red-sideband transition, i.e., $\Delta = 0$, the two resonators are parametrically coupled to each other.
Then, the excitation of the SAW resonator is converted to the excitation of the MW resonator, and its output MW power $P_\mathrm{out}$ can be written as
\begin{eqnarray}
P_\mathrm{out} &=& \hbar \omega _\mathrm{m}\kappa _\mathrm{ex}n_\mathrm{s}\frac{4 C_0 n_\mathrm{d}}{(1+ C_0 n_\mathrm{d})^2},
\end{eqnarray}
where 
$C_0\equiv 4g_0^2/(\kappa\Gamma)$ is the single-photon cooperativity between the SAW and MW resonators, $\kappa$ and $\Gamma$ are the respective total loss rates, and $\kappa_\mathrm{ex}$ is the external coupling of the MW resonator.
The average intra-resonator photon number $n_\mathrm{d}$ of the drive field and the intra-resonator phonon number $n_\mathrm{s}$ of the SAW resonator are calibrated by the saturation effect and the Stark shift of the MW resonator, respectively (See Supplementary Figures~4 and 5 in the Supplementary Note 5 and 6~\cite{supple}). 
Here, for the phonon-to-photon conversion, we use a weak drive field which provides a small $n_\mathrm{d}$~($\sim 0.01$) to avoid saturating the MW resonator. 
Therefore, we can evaluate $C_0$ from Eq.~(8). Acccording to the definition of $g_0$ in Eq.~(6), the relation between $C_0$ and $\beta$ follows
\begin{equation}
C_0=\frac{16g_\mathrm{p}^2\beta^2}{\kappa\Gamma (\omega _\mathrm{m}-\omega _\mathrm{s})^2},
\end{equation}
from which we evaluate the strength of the Pockels nonlinearity $\beta ~[=(\sqrt{C_0 \kappa \Gamma}/4g)(\omega_\mathrm{m} - \omega_\mathrm{s})]$ shown in Fig.~2b.
The overall behavior agrees well with the theoretical prediction.

\begin{figure}[th]
\begin{center}
   \includegraphics[width=7.5cm,angle=0]{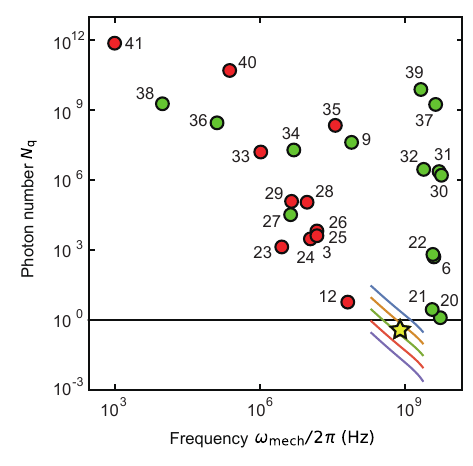}
\caption{
{\bf Single-photon quantum regime. }
Each data point shows intra-resonator drive photon number $N_\mathrm{q}$ necessary for the quantum-limited quadrature measurement of phonons as a function of mechanical eigenmode frequency~$\omega_\mathrm{mech} /2\pi$.
Green (red) circles indicate the values found in other opto-(electro-)mechanical systems with radiation pressure interaction. 
The yellow star shows the value obtained in  this work.
Color curves show the expectations from numerical simulations.
Different colors correspond to the values of $E_\mathrm{J}/E_\mathrm{C}= \{ 10^5, 10^4, 10^3, 10^2, 10 \}$ from top to bottom.
A list of the references indicated by the numbers next to the circles is presented in the reference list.
}
\label{fig4}
\end{center}
\end{figure}

{\bf Single photon quantum regime.}
We apply a stronger red-sideband drive to obtain a larger cooperativity with the artificial radiation pressure.
This results in the increase of the effective decay rate of the SAW resonator through the optomechanical damping.
Figure~\ref{fig3}a shows the spectra of the SAW resonator in the presence of the optomechanical damping rate $\Gamma _\mathrm{opt}$.
The total linewidth $\Gamma_\mathrm{all}$ of the spectrum is given by
\begin{equation}
\Gamma _\mathrm{all}=\Gamma + \Gamma _\mathrm{opt}=(1+C)\Gamma ,
\end{equation}
from which we evaluate the cooperativity $C$.

Figure \ref{fig3}b shows the cooperativity $C$ as a function of the drive photon number $n_\mathrm{d}$ at $\Phi = 0.445 \Phi_0 = \Phi_{\alpha=0}$ and $\Phi = 0.264 \Phi_0$, respectively.
Because of the absence of the self-Kerr nonlinearity, the saturation effect is much less pronounced at $\Phi = \Phi_{\alpha=0}$, allowing us to drive the system, as in other conventional optomechanical systems, to enhance the radiation pressure interaction.  
The cooperativity reaches 5 at high drive power. 
The remaining saturation effect is presumably due to the higher-order nonlinearities beyond the third order. 
The slope of the cooperativity for the small drive photon number corresponds to the single-photon cooperativity $C_0$, which is determined to be $1.7 \pm 0.1$ from the linear fit in Fig.~\ref{fig3}.
Thus, the single-photon quantum regime $C_0>1$ is achieved here.
From this value, the optomechanical coupling strength is evaluated to be $g_0/2\pi= 190$~kHz, which agrees well with the calculation shown in Fig.~\ref{fig1}g. 
It is also consistent with the estimation from the peak power of the up-conversion signal~[Eq. (9)], which gives $g_0/2\pi =230$~kHz.  


\section{DISCUSSIONS}
Having established the single-photon quantum regime, we compare various realizations of optomechanical systems from the viewpoint of quantum-limited measurement of phonons.
Figure \ref{fig4} shows the minimum intra-resonator photon number $N_\mathrm{q}=(n_\mathrm{th}+1/2)/C_0$, whose back-action shot noise on the mechanical mode becomes dominant over the thermal and vacuum noises~\cite{Kippenberg2014}.
Here $n_\mathrm{th} = 1/(e^{\hbar \omega_\mathrm{mech} /k_\mathrm{B} T} -1)$, $\omega_\mathrm{mech}$ is the mechanical mode frequency, $T$ is the bath temperature, and $k_\mathrm{B}$ is Boltzmann constant. 
The experiment was conducted at $T= 40$~mK, which gives $N_\mathrm{q}$ of $0.67\pm 0.04$. 
$N_q$ becomes less than unity when the kick by a single intra-resonator photon is larger than the mechanical oscillator beyond its noise amplitude of the motion.
In this regime, the quantum-limited quadrature measurement of the mechanical oscillator and the mechanically induced transparency of the MW resonator can be realized with single intra-resonator photons.
The current device is in the resolved-sideband limit, however, such that the incident drive is strongly filtered by the resonator and needs to be a large amplitude even if the required intra-resonator photon number is less than unity. 
To solve the problem, we can use a triple resonance technique~\cite{triple} to improve the coupling efficiency of the incident drive and utilize a quantum feature of the drive field.

The single-photon quantum regime is also useful for the quantum control of the mechanical oscillator through the radiation pressure interaction.
When $N_\mathrm{q}$ is less than unity, we can safely use a two-level system, i.e., a qubit, instead of a nearly harmonic microwave resonator. Then, quantum control of the SAW resonator can be readily demonstrated.
Recently, quanta of acoustic waves are being controlled and monitored by using a resonantly \cite{cleland2018, schoe2018} or dispersively \cite{lehnert2019, Amir2019} coupled superconducting qubit.
In contrast to those demonstrations, the radiation pressure interaction between a SAW resonator and a superconducting qubit gives rise to a `spin' dependent force, which can be used, e.g., for generating a Sch\"{o}dinger's cat state of a SAW resonator, as recently demonstrated in a vibrational mode in a trapped-ion experiment~\cite{homes}.
The spin-dependent force also enables fast entangling gates in analogy with those in trapped-ion systems~\cite{Wineland, Lucas}.

Moreover, as the numerical simulations indicate~(Color lines in Fig.~4), we can readily increase the coupling strength further by using a SAW resonator with higher frequency to reduce the detuning from the MW resonator.
The condition $g_0/\kappa >1$ is also within the scope of the future experiment, where the presence of a single phonon shifts the resonance frequency of the MW resonator by more than its linewidth~\cite{kurn}.
It will then allow for observation of quantum jumps between phonon Fock states~\cite{fock}.

\section*{Method}
{\bf Sample parameters.}
The parameters in the sample are the following.
At zero flux bias, the resonance frequency of the MW resonator is $\omega _\mathrm{m}/2\pi = 5.98$~GHz, and the internal and total loss rates are 10~MHz and 55~MHz, respectively.
At $\Phi = \Phi_{\alpha=0} \equiv 0.445 \Phi_0$, the resonance frequency is 3.85 GHz, the internal loss rate $\kappa_\mathrm{in} /2\pi = 3$ MHz, and the external loss rate $\kappa_\mathrm{ex} /2\pi = 17$ MHz. 
The resonance frequency of the SAW resonator is found to be $\omega _\mathrm{s}/2\pi = 785.25$~MHz, together with the total loss rate $\Gamma /2\pi =4.4$~kHz. The external coupling rate is designed to be $\Gamma _\mathrm{ex}/2\pi =0.6$~kHz.
The total loss rate of the MW resonator $\kappa /2\pi = 20$~MHz is much smaller than the resonance frequency of the SAW resonator. 
The piezoelectric coupling strength evaluated from the frequency shift of the SAW resonator under a strong drive is $g_\mathrm{p}/2\pi =6.4\mathrm{~MHz}$~\cite{noguchiSAW}.
The strength of the self-Kerr nonlinearity at zero flux bias is determined as $\alpha _0 /2\pi = -13.0$~MHz (See the details in the Supplementary Note 5~\cite{supple}).
The bath temperature in the experiment was at $T= 40$~mK.

\section*{Data availability}
The data that support the findings of this study are available from the corresponding
author upon reasonable request.

\section*{Acknowledgements}
The authors acknowledge K. Kusuyama for the help in sample fabrication. 
This work was partly supported by JSPS KAKENHI (Grant Number 26220601), JST PRESTO (Grant Number JPMJPR1429), and JST ERATO (Grant Number JPMJER1601).

\section*{Author contributions}
AN designed and fabricated the devices.
AN also conducted the experiments and analyzed the data.
YN supervised the project.
AN, RY, YT and YN discussed the results and wrote the manuscript together.

\section*{Competing interests}
The authors declare no competing interests.


\pagebreak

\begin{center}
\textbf{\Large Supplementary information for \\ Single-photon quantum regime of artificial radiation pressure \\ on a surface acoustic wave resonator}
\end{center}

\renewcommand{\figurename}{Supplementary Figure}
\renewcommand{\theequation}{S\arabic{equation}}
\setcounter{figure}{0}
\setcounter{equation}{0}

\section{Supplementary Note 1: Sample}
The circuit was fabricated on a 500-$\mu$m-thick ST-X cut quartz substrate. 
The Bragg mirrors, the interdigitated transducers (IDTs), the nonlinear microwave~(MW) resonator, and the coplanar waveguides for the external feed lines were simultaneously patterned in a wet-etching process from a 50-nm-thick evaporated aluminum film.
The Bragg mirrors have 750 fingers each. 
The IDT for the external coupling has a pair of four fingers, and the IDT connected to the MW resonator has a pair of ten fingers~(Fig.~1e in the main text). 
All those fingers have a width and a spacing of 1 $\mathrm{\mu m}$. 
The length of the surface-acoustic-wave~(SAW) resonator, the inner distance between the Bragg mirrors, is 240 $\mathrm{\mu m}$. 
The widths of the Bragg mirrors and the IDTs are 500~$\mu$m. 
The Josephson junctions for the SNAIL are made from Al/AlO$_x$/Al junctions, which are
 simultaneously fabricated by the shadow evaporation technique with the bridgeless resist mask. 
The size of the junctions are 150 $\times$ 150 nm for the small one and 300 $\times$ 300 nm for the large ones.

\begin{figure*}[bt]
\begin{center}
   \includegraphics[width=11cm,angle=0]{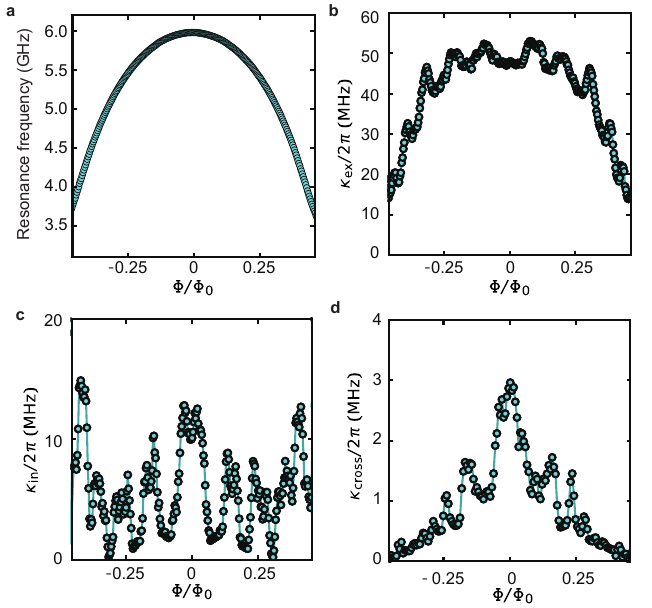}
\caption{
Parameters of the nonlinear MW resonator as a function of the flux bias.
{\bf{a.}}~The resonance frequency $\omega_\mathrm{m}/2\pi$, {\bf{b.}}~external loss rate $\kappa_\mathrm{ex}$, {\bf{c.}}~internal loss rate $\kappa_\mathrm{in}$ and {\bf{d.}} acoustic external loss rate to the IDT electrode are evaluated from the spectroscopy at the low-power limit with input-output theory.
}
\label{figs2}
\end{center}
\end{figure*}

Supplementary Figure \ref{figs2} shows the resonance frequency and the loss rates of the MW resonator as a function of the magnetic flux in the SNAIL loop.
Note that the loss rates are periodically fluctuating depending on the flux bias. 
The periodic modulation is presumably caused by the resonant acoustic radiation from the MW resonator.
The internal loss rate $\kappa _\mathrm{in}$ is divided into the electric loss $\kappa _\mathrm{e}$ and the acoustic radiation loss $\kappa _\mathrm{a}$ from the MW resonator.
The part of the acoustic radiation is picked up by the IDT electrode of the SAW resonator, and thus $\kappa _\mathrm{a}=\kappa _\mathrm{cross}+\kappa _\mathrm{rad}$, where $\kappa_\mathrm{cross}$ is the external coupling rate of the MW resonator though acoustic waves to the SAW input port (port 3 in Fig.~1b) and $\kappa _\mathrm{rad}$ is the acoustic radiation rate to the environment.
Supplementary Figure \ref{figs2}d shows the acoustic external coupling rate $\kappa _\mathrm{cross}$ of the MW resonator to the SAW input port (port 3 in Fig.~1b).


\section{Supplementary Note 2: NONLINEAR RESONATOR WITH SNAIL}
\begin{figure}[bt]
\begin{center}
   \includegraphics[width=8cm,angle=0]{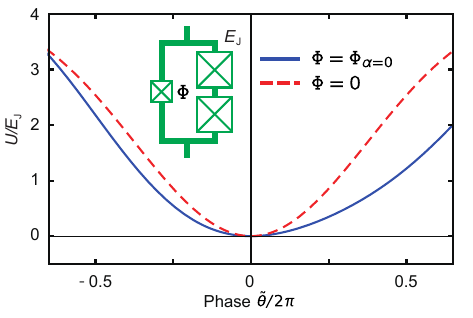}
\caption{
Inductive energy $U(\tilde{\theta} )$ of the SNAIL as a function of the phase difference $\tilde{\theta}$ across the small Josephson junction. The inset shows the circuit model of the SNAIL.
}
\label{figs3}
\end{center}
\end{figure}

Our SNAIL has a single small junction and two large junctions.
It is shunted with a large capacitor whose single-electron charging energy $E_\mathrm{C}$ is estimated to be $h\times 35$~MHz.
To determine the Josephson energies in the device, we fit the flux-dependent spectrum in Fig.~1g and obtain $E_\mathrm{J}^\prime = h \times 47.5 $~GHz and $E_\mathrm{J}= h \times 163.5 $~GHz, respectively. 
Supplementary Figure \ref{figs3} shows the inductive energy $U(\theta)$ of the SNAIL, given by Eq.(1) in the main text, in units $E_\mathrm{J}$.
For $\Phi \neq 0$, the parity symmetry is broken and the Pockels nonlinearity appears.
The inductive energy is expanded around the minimum at $\theta_0$ in a power series of $\tilde{\theta} \equiv \theta - \theta_0$ as
\begin{equation}
U(\tilde{\theta})/ E_\mathrm{J} =-\sum \chi _i \tilde{\theta}^i.
\end{equation}
The Hamiltonian of this nonlinear resonator in the transmon limit ($E_\mathrm{J}\gg E_\mathrm{C}$) reads
\begin{equation}
\hat{H}=4E_\mathrm{C} \hat{N}^2-E_\mathrm{J}\sum \chi _i \hat{\theta}^i,
\end{equation}
where $\hat{N}$ is the number operator of the excess Cooper pairs in the superconducting electrode connected to the ground via the SNAIL.
For the phase operator $\hat{\theta}$, we omit the tilde for simplicity. 
This Hamiltonian can be rewritten with the creation and annihilation operators as
\begin{eqnarray}
\hat{H}&=&\left(\sqrt{16E_\mathrm{C}E_\mathrm{J}\chi _2}-12E_\mathrm{C}\frac{\chi _4}{\chi _2}\right) \,\hat{a}^\dagger\hat{a}\nonumber\\
&&-3E_\mathrm{C}\left( \frac{\chi_2 E_\mathrm{J}}{E_\mathrm{C}}\right) ^{\! 1/4}\frac{\chi _3}{\chi _2} (\hat{a}^\dagger\hat{a}^\dagger\hat{a} + \mathrm{h.c.} ) \nonumber\\
&&-6E_\mathrm{C}\frac{\chi _4}{\chi _2}\hat{a}^\dagger\hat{a}^\dagger\hat{a}\hat{a}+O(\hat{a}^5),
\end{eqnarray}
where
\begin{equation}
\hat{a}=\left( \frac{E_\mathrm{C}}{h^2 \chi_2 E_\mathrm{J}} \right) ^{\! 1/4} \left( i\hat{N}+\sqrt{\frac{\chi_2 E_\mathrm{J}}{4E_\mathrm{C}}}\hat{\theta}\right) .
\end{equation}
The Pockels (second-order) and self-Kerr (third-order) nonlinearities appear in Eq.(S3). 
This relates the circuit parameters to the coefficients of the nonlinear tems in Eq.~(2) of the main text as 
\begin{eqnarray}
\alpha _0&=&-12E_\mathrm{C}\frac{\chi _4}{\chi _2},\\
\beta&=&-3E_\mathrm{C}\left( \frac{\chi_2 E_\mathrm{J}}{E_\mathrm{C}}\right) ^{\! 1/4}\frac{\chi _3}{\chi _2}.
\end{eqnarray}

\begin{figure*}[h]
\begin{center}
   \includegraphics[width=16cm,angle=0]{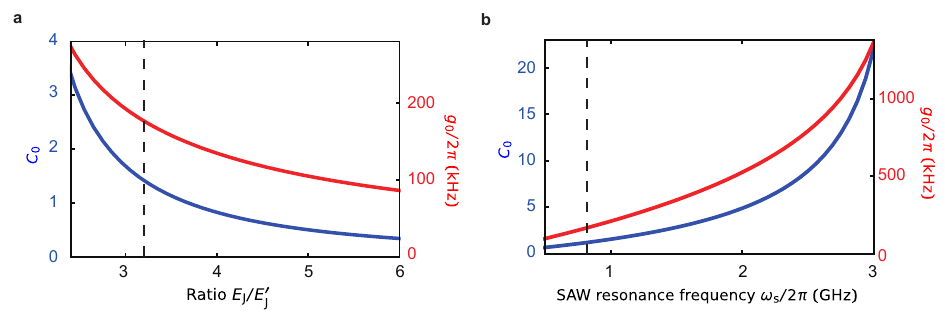}
\caption{
Estimation of the optomechanical coupling strength.
Blue and red curves show the single-photon cooperativity $C_0$ and the strength of the single-photon radiation pressure interaction $g_0$, respectively.
The Josephson energy $E_\mathrm{J}$ of the larger junctions in the SNAIL and the SAW resonator frequency $\omega_\mathrm{s}/2\pi$ are swept in {\bf{a}} and {\bf{b}}, respectively.
Vertical dashed lines indicate the parameters in the current experiment.
Other parameters are set to the values obtained in the experiment.
}
\label{figs4}
\end{center}
\end{figure*}

\section{Supplementary Note 3: Artificial optomechanical coupling}
The total Hamiltonian of the hybrid system consisting of a nonlinear MW resonator and a SAW resonator piezoelectrically coupled to each other is described without rotating wave approximation as 
\begin{eqnarray}
\hat{H}=\hat{H}_0+\hat{V}_0,
\end{eqnarray}
where
\begin{equation}
\hat{H}_0=\omega _\mathrm{m}\hat{a}^\dagger\hat{a}+\omega _\mathrm{s}\hat{b}^\dagger\hat{b},
\end{equation}
and
\begin{eqnarray}
\hat{V}_0&=&\beta (\hat{a}^\dagger\hat{a}^\dagger\hat{a} + \mathrm{h.c.} )
+\alpha _0 \hat{a}^\dagger\hat{a}^\dagger\hat{a}\hat{a},\nonumber\\
&&+g_\mathrm{p}(\hat{a}^\dagger +\hat{a})(\hat{b}^\dagger +\hat{b}).
\end{eqnarray}
The parameters and the operators are defined in the main text.
$g_\mathrm{p}$ is the piezoelectric coupling coefficient of the nonlinear MW resonator and SAW resonator.

By treating $\hat{V}_0$ as a perturbation, we find an effective Hamiltonian as
\begin{eqnarray}
\hat{H}_\mathrm{eff}&=&\hat{H}_0+\left( \alpha _0 -\frac{3\beta ^2}{\omega _\mathrm{m}}\right) \hat{a}^\dagger\hat{a}^\dagger\hat{a}\hat{a}
\nonumber\\
&&-\left( \frac{g_\mathrm{p}\beta}{\delta}+\frac{g_\mathrm{p}\beta}{\omega _\mathrm{m}}\right) \hat{a}^\dagger\hat{a}(\hat{b}^\dagger +\hat{b})\nonumber\\
&&-\frac{g_\mathrm{p}\beta\omega _\mathrm{s}}{2\omega _\mathrm{m}\delta}(\hat{a}^\dagger\hat{a}^\dagger\hat{b}+\mathrm{h.c.})\nonumber\\
&&+\frac{g_\mathrm{p}\beta\omega _\mathrm{s}}{2\omega _\mathrm{m}(\omega _\mathrm{m}+\omega _\mathrm{s})}(\hat{a}^\dagger\hat{a}^\dagger\hat{b}^\dagger+\mathrm{h.c.})\nonumber\\
&&-\frac{2g_\mathrm{p}\alpha _0}{\omega _\mathrm{m}+\omega _\mathrm{2}}(\hat{a}^\dagger\hat{a}\hat{a}\hat{b}+\mathrm{h.c.})\nonumber\\
&&-\frac{2g_\mathrm{p}\alpha _0}{\omega _\mathrm{m}-\omega _\mathrm{2}}(\hat{a}^\dagger\hat{a}\hat{a}\hat{b}^\dagger +\mathrm{h.c.})\nonumber\\
&&-\frac{2\alpha _0\beta}{\omega _\mathrm{m}}(\hat{a}^\dagger\hat{a}^\dagger\hat{a}\hat{a}\hat{a}+\mathrm{h.c.}),
\end{eqnarray}
where $\delta =\omega _\mathrm{m}-\omega_\mathrm{s}$ is the detuning between the MW and SAW resonators.
This calculation is valid when  $\{\omega _\mathrm{m},~\omega_\mathrm{s}, \delta\}\gg \{|\alpha _0|,~|\beta|,~|g_\mathrm{p}|\}$ is satisfied.
While the second term on the right-hand side gives the self-Kerr nonlinearity, the third term leads to the radiation pressure interaction, and the fourth term introduces dynamical Casimir effect.
When $\alpha _0=3\beta ^2/\omega _\mathrm{m}$, the self-Kerr nonlinearity vanishes, and the effective Hamiltonian is rewritten as
\begin{eqnarray}
\hat{H}_\mathrm{eff}&=&\hat{H}_0-2\frac{g_\mathrm{p}\beta}{\delta}\hat{a}^\dagger\hat{a}(\hat{b}^\dagger +\hat{b}),\\
&=&\hat{H}_0+g_0\hat{a}^\dagger\hat{a}(\hat{b}^\dagger +\hat{b}),
\end{eqnarray}
with the rotating wave approximation and a large detuning ($\delta\sim\omega _\mathrm{m}$).

\begin{figure}[h]
\begin{center}
   \includegraphics[width=8.0cm,angle=0]{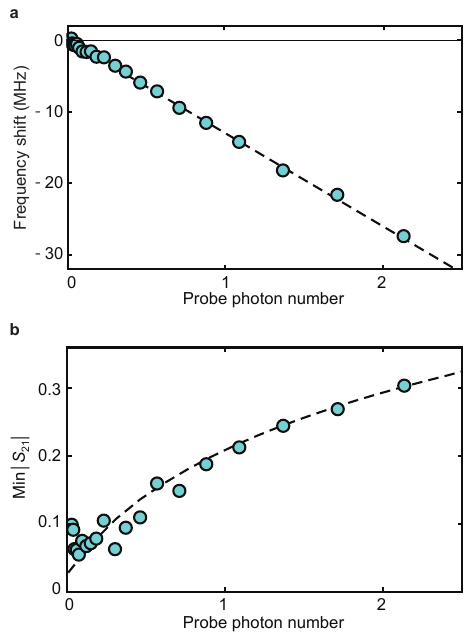}
\caption{Calibration of the resonator photon number and the nonlinearity of the MW resonator at zero flux bias.
{\bf{a.}}~Frequency shift of the nonlinear MW resonator as a function of the probe power represented by the intra-resonator photon number.
{\bf{b.}}~Saturation of the nonlinear MW resonator as a function of the probe power. 
The vertical axis shows the minimum values of the normalized transmission coefficient $|S_{21}|$ at the resonance of the MW resonator.
The self-Kerr nonlinearity makes the MW resonator saturated. 
}
\label{figs6}
\end{center}
\end{figure}

\section{Supplementary Note 4: Linearized Hamiltonian}
We irradiate the MW drive at frequency $\omega _\mathrm{d}$, the annihilation operator of the MW resonator becomes
\begin{equation}
\hat{a}\rightarrow e^{-i\omega _\mathrm{d}t}\Omega +\hat{a}
\end{equation}
and the interaction term becomes
\begin{eqnarray}
\hat{V}&=&g_0\hat{a}^\dagger\hat{a}(\hat{b}^\dagger +\hat{b})\\
&\rightarrow &g_0(e^{-i\omega _\mathrm{d}t}\Omega \hat{a}^\dagger+e^{i\omega _\mathrm{d}t}\Omega ^*\hat{a})(\hat{b}^\dagger +\hat{b})\nonumber \\
&&+g_0(|\Omega |^2 + \hat{a}^\dagger\hat{a}) (\hat{b}^\dagger +\hat{b}),
\end{eqnarray}
where $\Omega$ is the complex amplitude of the MW drive.

On the rotating frame with a unitary operator
\begin{equation}
\exp{[-i(\omega _\mathrm{d}+\omega _\mathrm{s})t\hat{a}^\dagger\hat{a}-i\omega _\mathrm{s}t\hat{b}^\dagger\hat{b}]}, 
\end{equation}
the Hamiltonian becomes
\begin{eqnarray}
\hat{H}_\mathrm{eff}=&&(\omega _\mathrm{m}-\omega _\mathrm{d}-\omega _\mathrm{s})\hat{a}^\dagger\hat{a}\nonumber \\
&&+g_0(\Omega \hat{a}^\dagger +\Omega ^*\hat{a})(\hat{b}^\dagger +\hat{b})\nonumber \\
&&+g_0(|\Omega |^2+\hat{a}^\dagger\hat{a}) (e^{i\omega_\mathrm{s}t}\hat{b}^\dagger +e^{-i\omega_\mathrm{s}t}\hat{b}).
\end{eqnarray}

When the bandwidth of the MW resonator $\kappa$ and the strength of the radiation pressure interaction $g_0$ are both smaller than $\omega _\mathrm{s}$, we can apply the rotating approximation to eliminate the third term and obtain the linearized Hamiltonian.

\section{Supplementary Note 5: Self-Kerr nonlinearity}
We characterize the amount of the self-Kerr nonlinearity of the MW resonator by measuring the frequency shift as a function of the probe power.
Supplementary Figures \ref{figs6}a and \ref{figs6}b show the frequency shift and the saturation of the absorption in the MW resonator at zero flux bias, respectively.
To analyze the result, we solve the master equation of the resonator with the third-order nonlinearity and fit the experimental data.
In the steady state, it fulfills
\begin{equation}
i[\hat{\rho} ,\hat{H}_\mathrm{fit}]+\hat{L}[ \hat{\rho} ] =\dot{\hat{\rho}}=0,
\end{equation}
where $\hat{\rho}$ is the density-matrix operator of the MW resonator, $\hat{L}$ is the Lindblad superoperator, and
\begin{eqnarray}
\hat{H}_\mathrm{fit}&=&\sqrt{4A_\mathrm{m}P_\mathrm{m}\kappa_\mathrm{ex}/\hbar\omega _\mathrm{m}} (\hat{a}^\dagger +\hat{a}) \nonumber\\
&&+ \Delta \hat{a}^\dagger\hat{a}+\alpha _0 \hat{a}^\dagger\hat{a}^\dagger\hat{a}\hat{a}.
\end{eqnarray}
Here, $A_\mathrm{m}$ is the attenuation through the input line of the MW feedline, and $P_\mathrm{m}$ is the probe power at the input port outside the refrigerator.
The saturation effect is highly nonlinear so that we can calibrate the absolute internal photon number with respect to the applied MW power. 
The strength of the self-Kerr nonlinearity and the attenuation in the input lines are determined from the fits as $\alpha_0/2\pi = -13.0$~MHz and $-57.3$~dB, respectively.

\section{Supplementary Note 6: Stark shift by the SAW excitation}
\begin{figure}[t]
\begin{center}
   \includegraphics[width=8cm,angle=0]{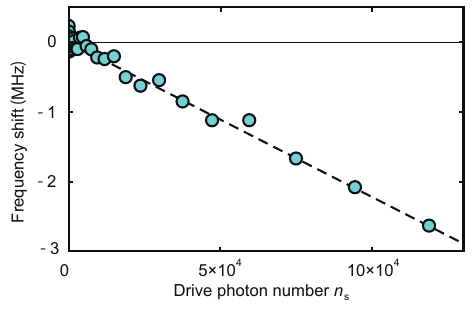}
\caption{
Stark shift of the MW resonator due to the excitation of phonons in the SAW resonator.
}
\label{figs5}
\end{center}
\end{figure}

To calibrate the SAW input power, we measure the Stark shift of the MW resonator induced by the SAW excitation.
Supplementary Figure \ref{figs5} shows the Stark shift as a function of the phonon number in the SAW resonator at zero flux bias.
Here we use the Stark shift per single phonon which is calculated to be
\begin{equation}
\chi _\mathrm{s}=\frac{2g_\mathrm{p}^2\alpha _0}{\delta ^2} = 2\pi \times 22\mathrm{~Hz}.
\end{equation}
The intra-resonator phonon number of the SAW resonator is given as $n_\mathrm{s}=4A_\mathrm{s}P_\mathrm{s}/\hbar\omega_\mathrm{s}\Gamma$, where $P_\mathrm{s}$ is the drive power at the SAW input port outside the refrigerator.
From comparison with the experimental result, the attenuation $A_\mathrm{s}$ along the SAW input line is determined to be $-73$~dB including the effect of the external coupling efficiency of the SAW resonator.

\begin{thebibliography}{99} 
\bibitem{Nichols} Nichols E. F. and Hull G. F., A preliminary communication on the pressure of heat and light radiation, \textit{Phys. Rev.} \textbf{13}, 307-320 (1901).
\bibitem{Kippenberg2014} Aspelmeyer M., Kippenberg T. J. and Marquardt F.,
Cavity optomechanics, \textit{Rev. Mod. Phys.} \textbf{86}, 1391-1452 (2014).

\bibitem{Teufel2011b} Teufel J. D., Donner T., Li D., Harlow J. W., Allman M. S., Cicak K., Sirois A. J., Whittaker J. D., Lehnert K. W. and Simmonds R. W., Sideband cooling of micromechanical motion to the quantum ground state, \textit{Nature} \textbf{475}, 359-363 (2011).
\bibitem{Teufel2015c} Lecocq F., Teufel J. D., Aumentado J. and Simmonds R. W.,
Resolving the vacuum fluctuations of an optomechanical system using an artificial atom, \textit{Nat. Phys.} \textbf{11}, 635-639 (2015).
\bibitem{noguchiMEM} Noguchi A., Yamazaki R., Ataka M., Fujita H., Tabuchi Y., Ishikawa T., Usami K. and Nakamura Y., 
Ground state cooling of a quantum electromechanical system with a silicon nitride membrane in a 3D loop-gap cavity,
\textit{New J. Phys.} \textbf{18}, 103036 (2016). 
\bibitem{Chan2011a} Chan J., Alegre T. P. M., Safavi-Naeini A. H., Hill J. T., Krause A., Gr\"{o}blacher S., Aspelmeyer M. and Painter O., Laser cooling of a nanomechanical oscillator into its quantum ground state, \textit{Nature} \textbf{478}, 89-92 (2011).
\bibitem{aspelmeyer} Hong S., Riedinger R., Marinkovic I., Wallucks A., Hofer S. G., Norte R. A., Aspelmeyer M., Gr\"{o}blacher S., 
Hanbury Brown and Twiss interferometry of single phonons from an optomechanical resonator, \textit{Science} \textbf{358}, 203-206 (2017).
\bibitem{Purdy} Purdy T. P., Grutter K. E., Srinivasan K. and Taylor J. M., 
Quantum correlations from a room-temperature optomechanical cavity, \textit{Science} \textbf{356}, 1265-1268 (2017).


\bibitem{Kippenberg2012} Verhagen E., Del\'{e}glise S., Weis S., Schliesser A. and Kippenberg T. J.,
Quantum-coherent coupling of a mechanical oscillator to an optical cavity mode, \textit{Nature} \textbf{482}, 63-67 (2012).

\bibitem{Sill2017} Santos J. T., Li J., Ilves J., Ockeloen-Korppi C. F. and Sillanp\"{a}\"{a} M., 
Optomechanical measurement of a millimeter-sized mechanical oscillator approaching the quantum ground state, \textit{New J. Phys.} \textbf{19}, 103014 (2017).


\bibitem{Painter2014}Meenehan S. M., Cohen J. D., Gr\"{o}blacher S., Hill J. T., Safavi-Naeini A. H., Aspelmeyer M. and Painter O., 
Silicon optomechanical crystal resonator at millikelvin temperatures,
\textit{Phys. Rev. A} \textbf{90}, 011803(R) (2014).
\bibitem{Sillanpaa2015} Pirkkalainen J.-M., Cho S. U., Massel F., Tuorila J., Heikkil\"{a} T. T., Hakonen P. J. and Sillanp\"{a}\"{a} M. A., 
Cavity optomechanics mediated by a quantum two-level system,
\textit{Nat. Commun.} \textbf{6}, 6981 (2015).

\bibitem{nakamura1999} Nakamura Y., Pashkin Y. A. and Tsai J. S., Coherent control of macroscopic quantum states in a single-Cooper-pair box, \textit{Nature} \textbf{398}, 786-788 (1999).
\bibitem{walraff2004nat}Wallraff A., Schuster D. I., Blais A., Frunzio L., Huang R.-S., Majer J., Kumar S., Girvin S. M. and Schoelkopf R. J., Strong coupling of a single photon to a superconducting qubit using circuit quantum electrodynamics, \textit{Nature} \textbf{431}, 162-167 (2004).
\bibitem{Yamamoto_JPA} Yamamoto T., Inomata K., Watanabe M., Matsuba K., Miyazaki T., Oliver W. D., Nakamura Y. and Tsai J. S., 
Flux-driven Josephson parametric amplifier, \textit{Appl. Phys. Lett.} \textbf{93}, 042510 (2008).
\bibitem{devoret2016} Roy A. and Devoret M. H.,
Introduction to quantum-limited parametric amplification of quantum signals with Josephson circuits,
\textit{C. R. Physique} \textbf{17}, 740-755 (2016).

\bibitem{noguchiSAW} Noguchi A., Yamazaki R., Tabuchi Y. and Nakamura Y., Qubit-assisted transduction for a detection of surface acoustic waves near the quantum limit, \textit{Phys. Rev. Lett.} \textbf{119}, 180505 (2017).

\bibitem{supple}\textit{Supplementary Information}

\bibitem{SNAIL} Frattini N. E., Vool U., Shankar S., Narla A., Sliwa K. M. and Devoret M. H., 3-wave mixing Josephson dipole element, 
\textit{Appl. Phys. Lett.} \textbf{110}, 222603 (2017).



\bibitem{1} Riedinger R., Hong S., Norte R. A., Slater J. A., Shang J., Krause A. G., Anant V., Aspelmeyer M. and Gr\"oblacher S., 
Non-classical correlations between single photons and phonons from a mechanical oscillator, 
\textit{Nature}, \textbf{530}, 313-316 (2016)

\bibitem{2} Meenehan S. M., Cohen J. D., Gr\"oblacher S., Hill J. T., Safavi-Naeini A. H., Aspelmeyer M. and Painter O., 
Silicon optomechanical crystal resonator at millikelvin temperatures, 
\textit{Phys. Rev. A} \textbf{90}, 011803(R) (2014)



\bibitem{5} Krause A. G., Hill J. T., Ludwig M., Safavi-Naeini A. H., Chan J., Marquardt F. and Painter O., 
Nonlinear Radiation Pressure Dynamics in an Optomechanical Crystal, 
\textit{Phys. Rev. Lett.} \textbf{115}, 233601 (2015)

\bibitem{6} Pirkkalainen, J.-M. and Damsk\"agg, E. and Brandt, M. and Massel, F. and Sillanp\"a\"a, M. A., 
Squeezing of Quantum Noise of Motion in a Micromechanical Resonator, 
\textit{Phys. Rev. Lett.} \textbf{115}, 243601 (2015)

\bibitem{7} Barzanjeh S., Redchenko E. S., Peruzzo M., Wulf M., Lewis D. P. and Fink J. M., 
Stationary Entangled Radiation from Micromechanical Motion, 
\textit{Nature} {\bf 570}, 480-483 (2019).


\bibitem{9} Lecocq F., Clark J. B., Simmonds R. W., Aumentado J. and Teufel J. D., 
Mechanically Mediated Microwave Frequency Conversion in the Quantum Regime, 
\textit{Phys. Rev. Lett.} \textbf{116}, 043601 (2016)

\bibitem{10} Lecocq F., Clark J. B., Simmonds R. W., Aumentado J. and Teufel J. D., 
Quantum Nondemolition Measurement of a Nonclassical State of a Massive Object, 
\textit{Phys. Rev. X} \textbf{5}, 041037 (2015)

\bibitem{11} Wilson, D. J., Sudhir, V., Piro, N., Schilling, R., Ghadimi, A. and Kippenberg, T. J., 
Measurement-based control of a mechanical oscillator at its thermal decoherence rate, 
\textit{Nature}, \textbf{524}, 325-329 (2015)

\bibitem{12} Teufel J. D., Lecocq F. and Simmonds R. W., 
Overwhelming Thermomechanical Motion with Microwave Radiation Pressure Shot Noise, 
\textit{Phys. Rev. Lett.} \textbf{116}, 013602 (2016)

\bibitem{13} Fink J. M., Kalaee M., Pitanti A., Norte R., Heinzle L., Davanco M., Srinivasan K. and Painter O., 
Quantum electromechanics on silicon nitride nanomembranes, 
\textit{Nat. Commun.} \textbf{7}, 12396 (2016)

\bibitem{14} Cohen J. D., Meenehan S. M., MacCabe G. S., Gr\"oblacher S., Safavi-Naeini A. H., Marsili F., Shaw M. D. and Painter O., 
Phonon counting and intensity interferometry of a nanomechanical resonator, 
\textit{Nature}, \textbf{520}, 522-525 (2015)

\bibitem{15} Fang K., Luo J., Metelmann A., Matheny M. H., Marquardt F., Clerk A. A. and Painter O., 
Generalized non-reciprocity in an optomechanical circuit via synthetic magnetism and reservoir engineering, 
\textit{Nat. Phys.} \textbf{13}, 465-471 (2017)

\bibitem{16} Balram K. C., Davanco M. I., Song J. D. and Srinivasan K., 
Coherent coupling between radiofrequency, optical and acoustic waves in piezo-optomechanical circuits, 
\textit{Nat. Phys.} \textbf{10}, 346-352 (2016)

\bibitem{17} Teufel J. D., Donner T., Castellanos-Beltran M. A., Harlow, J. W. and Lehnert, K. W., 
Nanomechanical motion measured with an imprecision below that at the standard quantum limit, 
\textit{Nat. Nano.} \textbf{4}, 820-823 (2009)

\bibitem{18} Shkarin A. B., Flowers-Jacobs N. E., Hoch S. W., Kashkanova A. D., Deutsch C., Reichel J. and Harris J. G. E., 
Optically Mediated Hybridization between Two Mechanical Modes, 
\textit{Phys. Rev. Lett.} \textbf{112}, 013602 (2014)


\bibitem{20} Singh V., Bosman S. J., Schneider B. H., Blanter Y. M., Castellanos-Gomez A. and Steele G. A., 
Optomechanical coupling between a multilayer graphene mechanical resonator and a superconducting microwave cavity, 
\textit{Nat. Nano.} \textbf{9}, 820-824 (2014)

\bibitem{21} Thompson J. D., Zwickl B. M., Jayich A. M., Marquardt F., Girvin S. M. and Harris J. G. E., 
Strong dispersive coupling of a high-finesse cavity to a micromechanical membrane, 
\textit{Nature}, \textbf{452}, 75 (2008)

\bibitem{22} Bochmann J., Vainsencher A., Awschalom D. D. and Cleland A. N., 
Nanomechanical coupling between microwave and optical photons, 
\textit{Nat. Phys.} \textbf{9}, 712-716 (2013)

\bibitem{23} Kleckner D., Pepper B., Jeffrey E., Sonin P., Thon S. M. and Bouwmeester D., 
Optomechanical trampoline resonators, 
\textit{Opt. Express.} \textbf{19}, 19708-19716 (2011) 

\bibitem{24} Mitchell M., Khanaliloo B., Lake D., Masuda T., Hadden J. and Barclay P., 
Single-crystal diamond low-dissipation cavity optomechanics, 
\textit{Optica},  \textbf{3}, 963-970 (2016). 

\bibitem{25} Regal C. A., Teufel J. D. and Lehnert K. W., 
Measuring nanomechanical motion with a microwave cavity interferometer, 
\textit{Nat. Phys.} \textbf{4}, 555-560 (2008)

\bibitem{26} Cuthbertson B. D., Tobar M. E., Ivanov E. N. and Blair D. G., 
Parametric back-action effects in a high-Q cryogenic sapphire transducer, 
\textit{Review of Scientific Instruments}, \textbf{67}, 2435-2442 (1996)



\bibitem{triple}Rueda A., Sedlmeir F., Collodo M. C., Vogl U., Stiller B., Schunk G., Strekalov D. V., Marquardt C., Fink J. M., Painter O., Leuchs G., and Schwefel H. G. L., 
Efficient microwave to optical photon conversion: an electro-optical realization,
\textit{Optica} \textbf{3}, 597-604 (2016)  

\bibitem{cleland2018}Satzinger K. J., Zhong Y. P., Chang H.-S., Peairs G. A., Bienfait A., Chou M.-H., Cleland A. Y., Conner C. R., Dumur \'{E}., Grebel J., Gutierrez I., November B. H., Povey R. G., Whiteley S. J., Awschalom D. D., Schuster D. I. and Cleland A. N., 
Quantum control of surface acoustic-wave phonons,
\textit{Nature} \textbf{563}, 661-665 (2018).

\bibitem{schoe2018}Chu Y., Kharel P., Yoon T., Frunzio L., Rakich P. T. and Schoelkopf, R. J.,
Creation and control of multi-phonon Fock states in a bulk acoustic-wave resonator,
\textit{Nature} \textbf{563}, 666-670 (2018).

\bibitem{lehnert2019} Sletten L. R., Moores B. A., Viennot J. J. and Lehnert K. W., 
Resolving phonon fock state in a multimode cavity with a double-slit qubit, 
\textit{Phys. Rev. X} \textbf{9}, 021056 (2019).

\bibitem{Amir2019} Arrangoiz-Arriola P., Wollack E. A., Wang Z., Pechal M., Jiang W., McKenna T. P., Witmer J. D., Safavi-Naeini A. H., 
Resolving the energy levels of a nanomechanical oscillator, 
\textit{Nature} \textbf{571}, 537–540 (2019).

\bibitem{homes} Kienzler D., Fl\"{u}hmann C., Negnevitsky V., Lo H.-Y., Marinelli M, Nadinger D. and Homes J. P., 
Observation of quantum interference between separated mechanical oscillator wave packets, 
\textit{Phys. Rev. Lett.} \textbf{116}, 140402 (2016).

\bibitem{Wineland}Turchette Q. A., Wood C. S., King B. E., Myatt C. J., Leibfried D., Itano W. M., Monroe C. and Wineland D. J., 
Deterministic entanglement of two trapped ions,
\textit{Phys. Rev. Lett.} \textbf{81}, 3631 (1998).

\bibitem{Lucas}Ballance C. J., Harty T. P., Linke N. M., Sepiol M. A. and Lucas D. M., 
High-fidelity quantum logic gates using trapped-ion hyperfine qubits,
\textit{Phys. Rev. Lett.} \textbf{117}, 060504 (2016).

\bibitem{kurn} Murch K. W., Moore K. L., Gupta S. and Stamper-Kurn D. M., 
Observation  of  quantum-measurement backaction with  an ultracold atomic gas, 
\textit{Nat. Phys.} \textbf{4}, 561-564 (2008).

\bibitem{fock}Miao H., Danilishin S., Corbitt T. and Chen Y., 
Standard quantum limit for probing mechanical energy quantization, 
\textit{Phys. Rev. Lett.} \textbf{103}, 100402 (2009).

\end{thebibliography}
\end{document}